\documentclass[11pt]{article}
\usepackage{geometry}
\usepackage{amsmath}
\usepackage{amssymb}
\usepackage{caption}
\usepackage{graphicx}
\usepackage{latexsym}
\RequirePackage{scrlfile}
\PreventPackageFromLoading{subfig}
\usepackage{times} 
\usepackage{subcaption}
\usepackage{svg}
\usepackage{hyperref}
\usepackage[english]{babel}
\usepackage{fancyhdr}
\usepackage{subfig}
\usepackage{setspace}
\usepackage{float}
\restylefloat{table}
\usepackage{array}
\usepackage{setspace}

\fancyfootoffset{1cm}
\fancypagestyle{plain}{%
\fancyhead{}
\fancyfoot{}
\rfoot{\thepage}
\lfoot{{\small EVS33}}

}

\renewenvironment{abstract}
 {
  {\bfseries \large{\abstractname}}
  \par
  \vspace{10pt}
  \normalsize
 }

\title
{\Large \emph{33\textsuperscript{nd} Electric Vehicle Symposium (EVS33)\\ Portland, Oregon, June 14-17, 2020}\\ \hspace{10pt}\\ \LARGE\bf 
On Energy Optimal Speed Trajectories: The Impact of Electric Powertrain Efficiency Characteristics}

\author{{\large            Arian Ahmadi$^1$, Peter H. Bauer$^2$}\\
        {\small $^1$\em Department of Electrical Engineering, University of Notre Dame, 275 Fitzpatrick Hall, Notre Dame, U.S.A.}\\
        {\small $^2$\em \{aahmadi, pbauer\}@nd.edu}}

\date{}
\begin{document}

\setlength{\parindent}{0mm}
\baselineskip 10pt

\maketitle
\renewcommand{\abstractname}{Summary}
\setstretch{1.25}
\rule{\textwidth}{1pt}
\begin{abstract}
This paper investigates the effect of the powertrain efficiency map on energy optimal speed trajectories, especially stop-to-stop trajectories. A variety of different efficiency maps are explored and the energy optimization process is carried out. It is shown that while the efficiency maps can have a significant impact
on the overall transportation energy, they have little impact on the shape of the optimal speed profiles.  A number of very different efficiency maps and their effects on speed trajectories are illustrated. Also, the impact that vehicle and segment parameters as well as average speeds have on the savings are investigated.  
This paper also presents the impact of transient penalty on the shape of the energy optimal trajectories and associated savings. It is observed that, by adding an additional transient penalty to the energy balance, the shape of trajectories changes and energy savings decrease. 
\\
\\
{\em Keywords: energy efficiency, speed trajectories, optimal, efficiency map, electric
vehicle}
\end{abstract}

\setstretch{1}
\rule{\textwidth}{1pt}

\section{Introduction}
Electric vehicles (EVs) are being actively developed by car companies
worldwide to achieve  higher energy efficiency and lower emissions.
However, the limited range of EVs can lead to range anxiety, which is one of
the causes of their slow market 
penetration. It is therefore of utmost importance to increase battery
capacity and range \cite{Fnoor17,  Zonggen14, Zonggen2018, Ahmadi,Zonggen15}. Similar to the work presented in this paper, \cite{Hooker88} investigates
the numerical optimization of speed trajectories for vehicles equipped with internal combustion
engines (ICEs) between stop
signs. In \cite{Henriksson17}, the authors study  energy-optimal speed profiles for a large number of
 ICEs vehicles without considering the
vehicle efficiency and its dependency on speed and torque. It has recently been shown in \cite{Eduardo19} that optimizing speed trajectories for stop-to-stop traffic can boost transportation efficiency significantly for electric drives. In the proposed algorithm, transportation energy is the only term in the cost function and all other conditions such as acceleration and average speed only appear as constraints. In that work, the authors used the combined efficiency of the motor and inverter of the Nissan Leaf as the basis for the general shape of the efficiency map.
In \cite{Eduardo20}, the authors explore vehicle-embedded algorithms that minimize energy usage in electric vehicles (EVs) using a lumped constant efficiency model instead of considering an efficiency map.
In other work, see \cite{Zonggen17}, the authors present the impact of environmental factors such as wind speed, rolling resistance and temperature on electric vehicle energy consumption. In \cite{Zonggen18}, a robust optimization model that exploits the aforementioned environmental factors to generate an optimal speed profile is derived.\\
This paper explores the effect of the different powertrain efficiency maps on energy optimal speed trajectories and energy savings in typical urban driving situations in stop-to-stop traffic. In order to obtain the optimal speed trajectories, we implement an algorithm that minimizes
energy usage in electric vehicles (EVs). The approach
taken can theoretically be applied to any type of vehicle, with different powertrain efficiency maps. The presented
concept minimizes the expended battery energy, given the
distance and the desired average speed between two stops.
Hence, the proposed method chooses the speed-versus-time profiles among infinitely many speed trajectories which satisfy the
given constraints and minimize transportation energy. \\
To tackle this type of optimization problem, we need to know the basic vehicle parameters, e.g., rolling resistance, air
drag coefficient, frontal cross-sectional area, vehicle mass, and powertrain efficiencies. The aforementioned parameters are well known for all electric vehicles. Using these predefined parameters, constraints and drive segment information, the proposed optimization algorithm creates the energy-optimal
speed profiles.\\
The rest of the paper is structured as follows: In Section 2, the energy and power flow model is introduced as well as the optimization problem with constraints. Section 3 provides three hypothetical efficiency maps as well as simulation results.
The transient analysis and its impact on the optimal trajectories are presented in Section 4. Conclusion is given in Section 5.

\section{Problem Formulation}
In order to optimize the expected energy of the battery, basic vehicle parameters for the formulation of the problem are needed. Hence, according to the models described in \cite{Zonggen17}, the wheel power, denoted as $P_{wheel}$ can be formulated as (\ref{system}) where $m$, $\Delta m$, $v(t)$, $\dot{v}(t)$, $C_d$ and $A$ are the vehicle’s mass, rotational equivalent mass, speed, acceleration, frontal drag coefficient and cross-sectional area, respectively. Additionally, $\rho$ is the air density which is 1.2$kg/m^3$ , $f_r$ is the coefficient of rolling resistance equal to 0.01 and $g$ is the gravitational acceleration which is 9.81$m/s^2$\cite{Ehsani18}. In the analysis presented, a flat surface is assumed.
\begin{equation}
P_{wheel}(t)= (m+\Delta m) v(t) \dot{v}(t)+\frac{1}{2} C_{d} A \rho v(t)^{3}+m g f_{r} v(t)
\label{system}
\end{equation}
The forward motion
and regenerative breaking power equation are given in (\ref{forward}) and (\ref{reverse}), respectively.

\begin{equation}
P_{bat}(t)=\frac{1}{\eta_{frw}(\omega,\text{T})}\left(P_{wheel}(t)\right) \quad \text { for } \quad P_{wheel}\geq0
\label{forward}
\end{equation}
\begin{equation}
P_{bat}(t)=\eta_{reg}(\omega,\text{T})\left(P_{wheel}(t)\right) \quad  \text { for } \quad P_{wheel}<0
\label{reverse}
\end{equation}
where $P_{bat}(t)$ is the power at the battery, $\eta_{frw}(\omega,\text{T})$ and $\eta_{reg}(\omega,\text{T})$ is the
vehicle’s efficiency for forward and reverse power flow, respectively. \text{T} is the torque of
the motor, and $\omega$ is the rotational speed of the motor. 
The battery energy, $E_{bat}$, is given by (\ref{E_bat}) where $t_f$ is the final time.

\begin{equation}
E_{bat}=\int_{0}^{t_f} P_{bat}(\tau) d\tau
\label{E_bat}
\end{equation}
The optimization problem can be written as shown in (\ref{optimization}).
Constraints are given by average speed $v_{avg}$, maximum acceleration and deceleration.
\begin{equation}
\begin{array}{ll}{\underset{v(t)}{\operatorname{minimize}}} & {\int_{0}^{t_{f}} P_{b a t}(\tau) d \tau} \\ {\text { s.t. }} & \frac{1}{t_f}{\int_{0}^{t_{f}} v(\tau) d \tau=v_{avg}} \\ {} & {\dot{v}_{\min }<\dot{v}(t)<\dot{v}_{\max }}\\&
{0 \leq v(t)\leq v_{max}}\end{array}
\label{optimization}
\end{equation}
where $P_{bat}(t)$ is given by (\ref{forward}) and (\ref{reverse}). The optimization problem implemented to generate the speed trajectories does not use jerk limitations (limitations on
the derivative of the acceleration) as a constraint.
The optimization problem can be formulated in a piecewise
discretized form, shown in (\ref{discrete1}), (\ref{discrete2}), and (\ref{discretefinal}). Note that, we approximated
the acceleration term with the difference in kinetic energy values at each distance
segment. Each discretized energy segment, $ E_{wheel}[n]$,
is described by (\ref{discrete1}) where $v_n$ is the discretized speed value at time $n\Delta t$ and $\Delta t$ the sampling time.

\begin{equation}
 E_{wheel}[n]=\frac{m}{2}\left(v_{n+1}^{2}-v_{n}^{2}\right)+\frac{1}{2} C_{d} A \rho v_{n}^{3} \Delta t+m g f_{r} v_{n} \Delta t
 \label{discrete1}
\end{equation}
The optimization problem in a piecewise discretized form, is shown in (\ref{discrete2}) and
(\ref{discretefinal}) where an energy segment is denoted by
$\Delta E_{bat}[n]$.

\begin{equation}
\Delta E_{bat}[n]=\left\{\begin{array}{ll}{\frac{1}{\eta_{frw}(\omega,\text{T})}  E_{wheel}[n]} & {\text { for } } \quad E_{wheel}[n] \geq 0 \\ {{\eta_{reg}(\omega,\text{T})} E_{wheel}[n]} & {\text { for } \quad E_{wheel}[n]<0}\end{array}\right.
\label{discrete2}
\end{equation}
The total energy used by the vehicle is given by $E_{bat}$, where $N$ represents the total number of time segments.
\begin{equation}
E_{bat}=\sum_{n=1}^{N} \Delta E_{bat}[n]
\label{discretefinal}
\end{equation}
Finally, the discrete optimization problem can be written as shown in (\ref{modelfinal}).

$$\min _{v_{n}} E_{bat}$$
\begin{equation}
\begin{cases}
     \text { s.t. } \sum_{n=1}^{N} \frac{v_{n}}{N}=v_{a v g}\\
     {0 \leq v_{n}\leq v_{max}}\\
     {d_{min} \leq \frac{v_{n+1}-v_{n}}{\Delta t}\leq a_{max}}, \quad  if \quad   n \in \{1,2,...,N-1\}\\[2\jot]
     {d_{min} \leq \frac{-v_{n}}{\Delta t}\leq a_{max}} \quad  if \quad\quad\quad   n = N\\[2\jot]
  \end{cases}
  \label{modelfinal}
\end{equation}
where $v_{avg}$ is the desired average speed, $v_{max}$ is the
maximum allowed speed, $a_{max}$ is the maximum acceleration and $d_{max}$ is the maximum deceleration.

\section{Simulations}
In order to characterize the segment between two successive stops, we need to know the distance $X$ and the average speed $v_{avg}$ between them. In this paper, the FTP-75 drive cycle was used as a baseline for typical traffic flow. In order to generate the typical traffic profile, the drive cycle was split into segments (stop-to-stop) which is within +/-\%20 of $X$, normalized in time and speed and then averaged over all segments. This approximation ensures that the acceleration is kept close to the typical acceleration in urban scenarios. The MATLAB $fmincon$ solver with the $Sequential$ $Quadratic$ $Programming$ $(SQP)$ algorithm is used to implemented the optimization problem expressed in (\ref{optimization}). $Fmincon$ is a nonlinear programming solver which is used to find the minimum of a constrained nonlinear multivariable function.\\
In section 3.2, a number of simulations were performed using different efficiency maps to analyze the impact of the powertrain efficiency map on energy optimal speed profiles . In these simulations the vehicle parameters as well as the trajectory parameters were varied. The electric vehicles Nissan Leaf and Tesla Model S were used as the
basis for the parameter sets. Table \ref{table:I} shows the parameter set
utilized for each vehicle in the simulations. Vehicle type 1
has a parameter set similar to the Nissan Leaf’s parameters
and Vehicle type 2 is similar to the Tesla Model S.\\
\begin{table}[H]
\centering
\caption{VEHICLE PARAMETERS UTILIZED IN SIMULATIONS}
\begin{tabular}{|l|c|c|c|c|c|}
\hline
        & Mass  & $C_dA$  & Max.   & Max.  \\
Vehicle &($kg$) & ($m^2$) & acceleration     & deceleration \\
        &       &         & $(m/s^2)$    & $(m/s^2)$       \\
\hline        
type 1 & 1525& 0.6583 & 4.6 & 2\\
\hline
type 2 & 2018& 0.6720 & 8 & 2.5\\
\hline
\end{tabular}
\label{table:I}
\end{table}
\subsection{Efficiency Maps}
\begin{figure}[H]
\centering
\subfloat[\label{main:Torque-dependent}]{
  \includegraphics[width=0.44\textwidth]{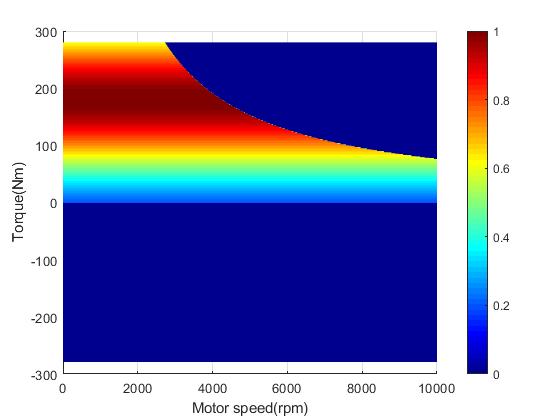}
  }
\subfloat[\label{main:ICE type}]{
  \includegraphics[width=0.4\textwidth]{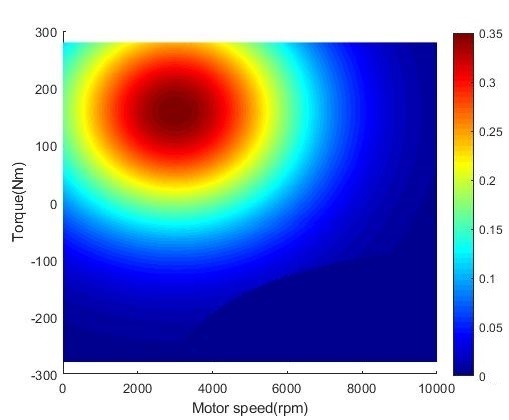}
  }\par
\subfloat[\label{main:Hybrid type}]{
  \includegraphics[width=0.43\textwidth]{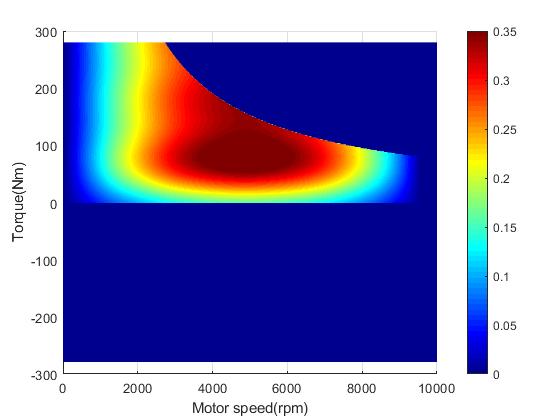}
  }
\subfloat[\label{main:Speed-dependent}]{
  \includegraphics[width=0.43\textwidth]{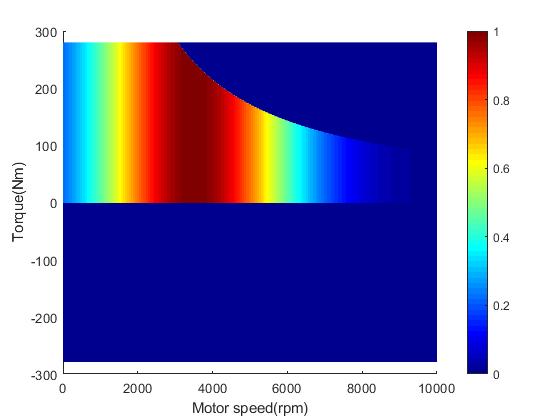}
  }
\end{figure}

\begin{figure}[H]
\centering
\subfloat[\label{main:Two high ridges}]{
  \includegraphics[width=0.43\textwidth]{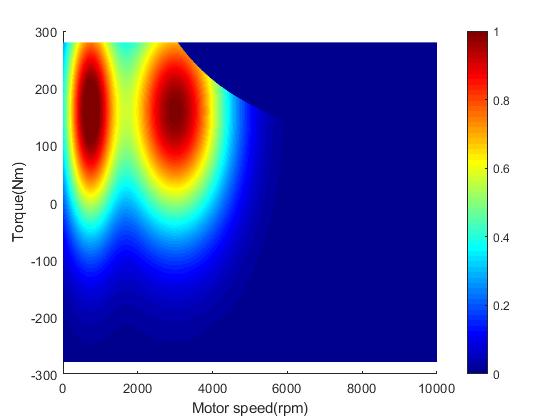}
  }\par
\caption[Hypothetical efficiency maps]{(a) PT efficiency characteristic -Type 1, (b) PT efficiency characteristic -Type 2, (c) PT efficiency characteristic -Type 3, (d) PT efficiency characteristic -Type 4 (e) PT efficiency characteristic -Type 5}
\label{fig:main}
\end{figure}

In order to describe the effect of the efficiency maps on energy optimal speed trajectories, five hypothetical maps were generated. Fig. (\ref{main:Torque-dependent}) shows the efficiency map of a vehicle with a highly efficient powertrain for high values of torque without any dependency on motor speeds. The second efficiency map, Fig. (\ref{main:ICE type}), represents a speed- and torque-dependent model and it is based on ICE type efficiency maps. Fig. (\ref{main:Hybrid type}) represents a Hybrid type efficiency map where the maximum efficiency from source to the wheel is quite small, around 0.35. Fig. (\ref{main:Speed-dependent}) represents the efficiency map only depends on speed of the motor without any dependency on motor torques. Finally, Fig (\ref{main:Two high ridges}) shows the efficiency map with two ridges of high efficiency.\\

All of the efficiency maps were generated
by multiplying a horizontal spline (ranging from zero to the
maximum speed of the electric motor) by a vertical spline
(ranging from the maximum negative to the maximum positive
torque of the motor). This approach was taken in order to
generate a continuously differentiable function $\eta(\omega,\text{T})$. 
\subsection{Simulation Results}
\begin{figure}[H]
\begin{minipage}{.32\linewidth}
\centering
\subfloat[]{\label{Torque-dependent:a}\includegraphics[scale=.28]{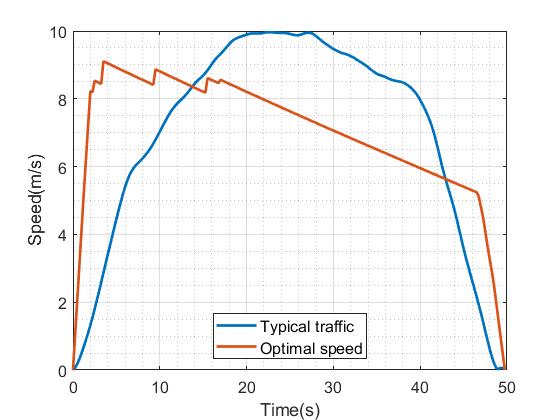}}
\end{minipage}
\begin{minipage}{.32\linewidth}
\centering
\subfloat[]{\label{Torque-dependent:b}\includegraphics[scale=.28]{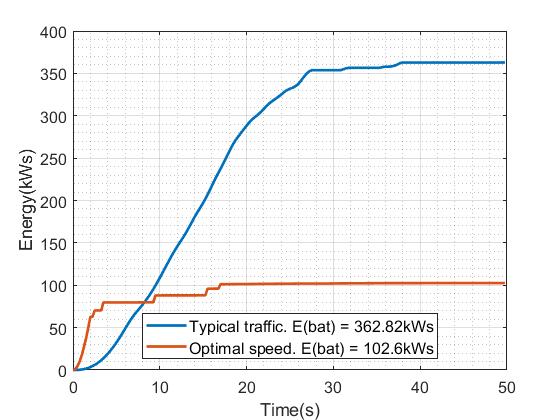}}
\end{minipage}
\begin{minipage}{.32\linewidth}
\centering
\subfloat[]{\label{Torque-dependent:c}\includegraphics[scale=.4]{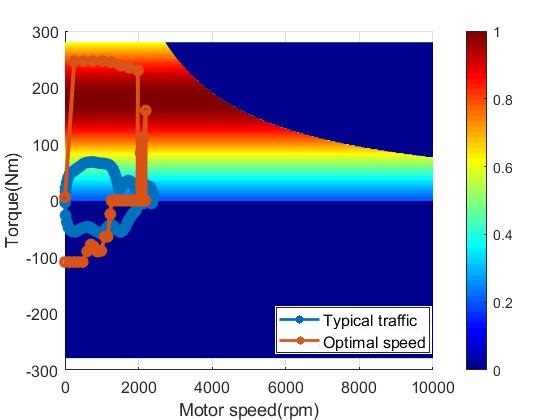}}
\end{minipage}
\caption[Two numerical solutions]{(a) Typical and optimal speed profiles, (b) cumulative energy consumption, (c) efficiency trajectories}
\label{fig:mainTorque-dependent}
\end{figure}

\begin{figure}[H]
\begin{minipage}{.32\linewidth}
\centering
\subfloat[]{\label{ICE type:a}\includegraphics[scale=.28]{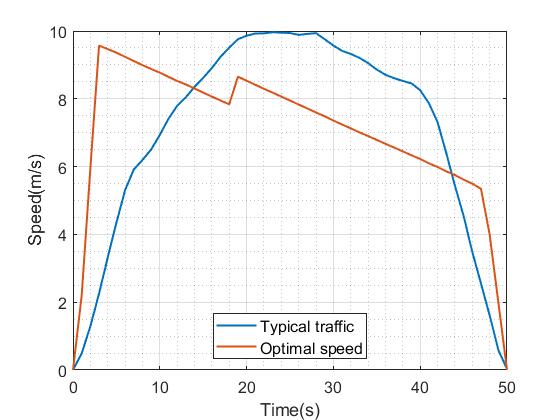}}
\end{minipage}
\begin{minipage}{.32\linewidth}
\centering
\subfloat[]{\label{ICE type:b}\includegraphics[scale=.28]{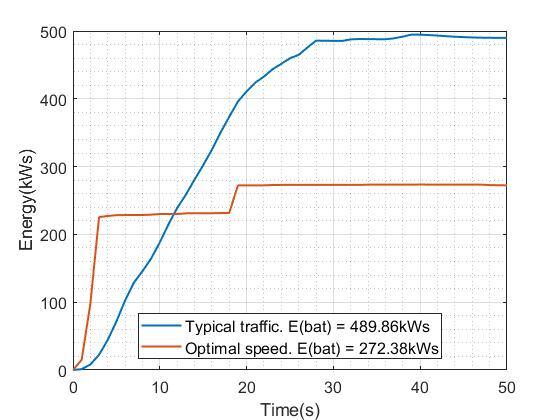}}
\end{minipage}
\begin{minipage}{.32\linewidth}
\centering
\subfloat[]{\label{ICE type:c}\includegraphics[scale=.4]{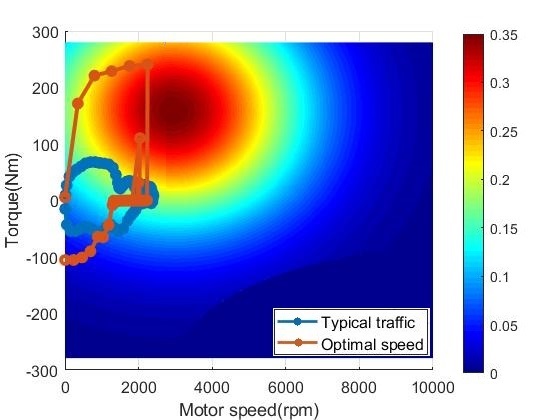}}
\end{minipage}

\caption[Two numerical solutions]{(a) Typical and optimal speed profiles, (b) cumulative energy consumption, (c) efficiency trajectories}
\label{fig:mainICE type}
\end{figure}

\begin{figure}[H]
\begin{minipage}{.32\linewidth}
\centering
\subfloat[]{\label{Hybrid type:a}\includegraphics[scale=.38]{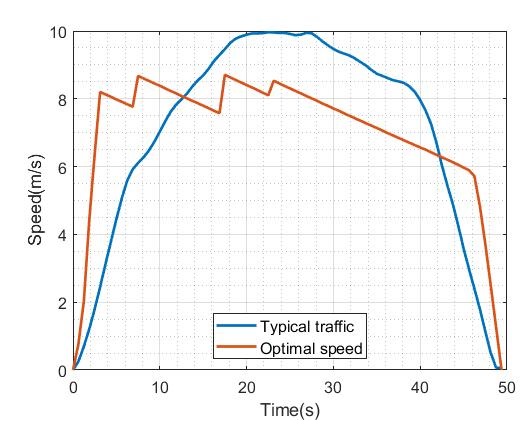}}
\end{minipage}
\begin{minipage}{.32\linewidth}
\centering
\subfloat[]{\label{Hybrid type:b}\includegraphics[scale=.38]{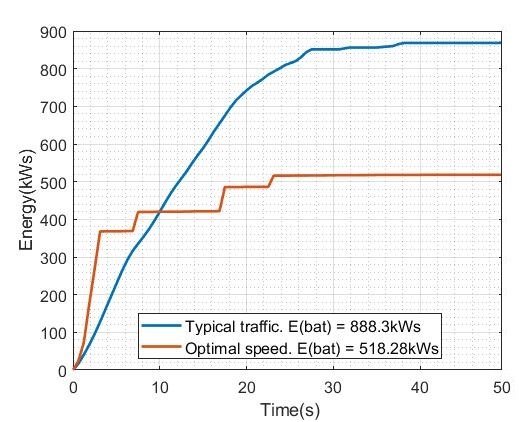}}
\end{minipage}
\begin{minipage}{.32\linewidth}
\centering
\subfloat[]{\label{Hybrid type:c}\includegraphics[scale=.4]{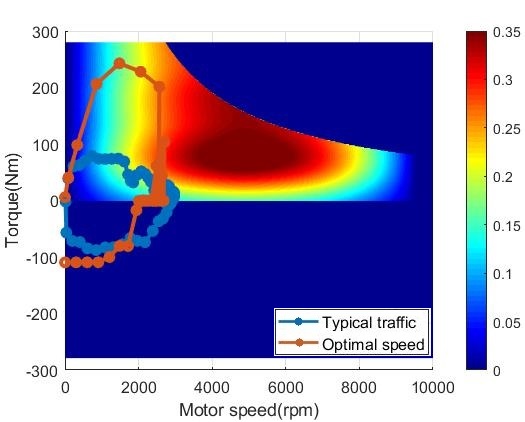}}
\end{minipage}

\caption[Two numerical solutions]{(a) Typical and optimal speed profiles, (b) cumulative energy consumption, (c) efficiency trajectories}
\label{fig:mainHybrid type}
\end{figure}

\begin{figure}[H]
\begin{minipage}{.32\linewidth}
\centering
\subfloat[]{\label{Speed-dependent:a}\includegraphics[scale=.28]{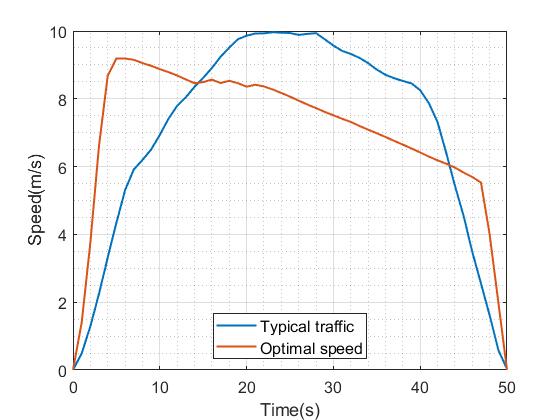}}
\end{minipage}
\begin{minipage}{.32\linewidth}
\centering
\subfloat[]{\label{Speed-dependent:b}\includegraphics[scale=.28]{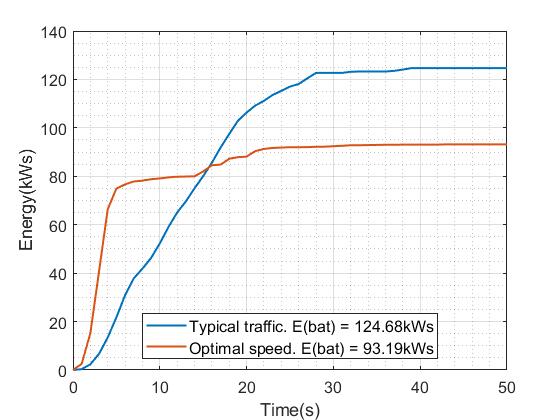}}
\end{minipage}
\begin{minipage}{.32\linewidth}
\centering
\subfloat[]{\label{Speed-dependent:c}\includegraphics[scale=.4]{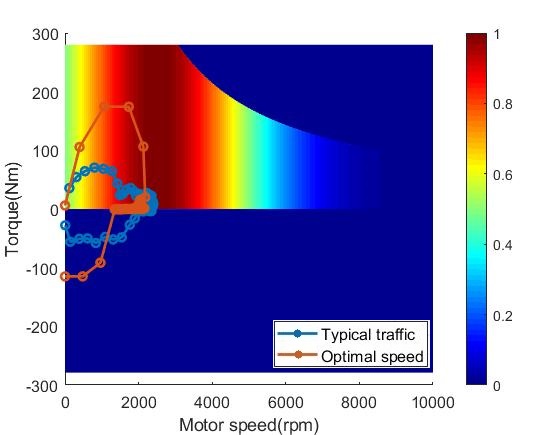}}
\end{minipage}

\caption[Two numerical solutions]{(a) Typical and optimal speed profiles, (b) cumulative energy consumption, (c) efficiency trajectories}
\label{fig:mainSpeed-dependent}
\end{figure}

\begin{figure}[H]
\begin{minipage}{.32\linewidth}
\centering
\subfloat[]{\label{Two high ridges:a}\includegraphics[scale=.28]{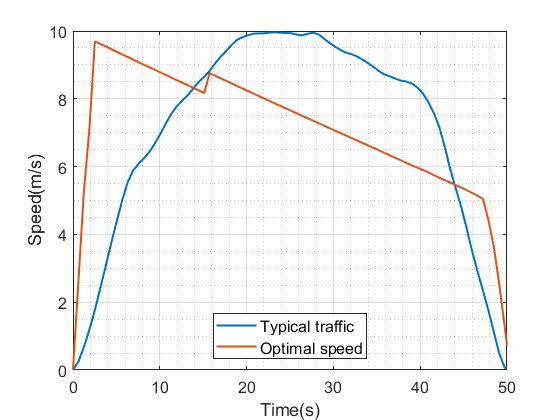}}
\end{minipage}
\begin{minipage}{.32\linewidth}
\centering
\subfloat[]{\label{Two high ridges:b}\includegraphics[scale=.28]{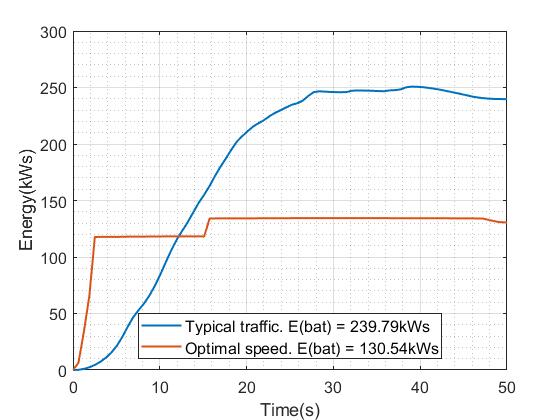}}
\end{minipage}
\begin{minipage}{.32\linewidth}
\centering
\subfloat[]{\label{Two high ridges:c}\includegraphics[scale=.39]{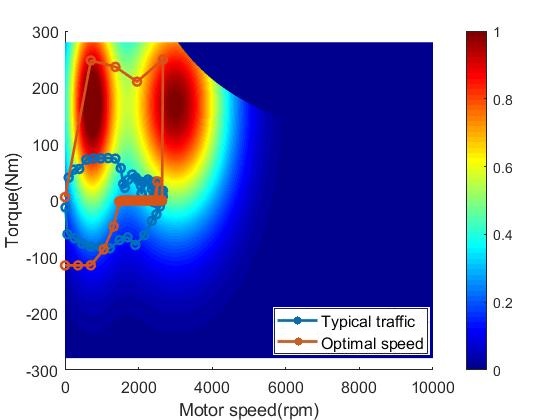}}
\end{minipage}

\caption[Two numerical solutions]{(a) Typical and optimal speed profiles, (b) cumulative energy consumption, (c) efficiency trajectories}
\label{fig:mainTwo high ridges}
\end{figure}

The results for a Nissan Leaf in a 350m segment based on the above-mentioned efficiency maps are shown in Fig. \ref{fig:mainTorque-dependent} to Fig. \ref{fig:mainTwo high ridges}, respectively.

Fig. \ref{fig:mainTorque-dependent} shows that the optimized speed profile consumes 102.6$kWs$ while the typical traffic baseline uses 362.82$kWs$, a 71.72\% reduction. This huge saving amount achieved because the optimizer operates in the highest efficiency region during first hard acceleration. \\
Utilizing the second efficiency maps previously
described, Fig. (\ref{main:ICE type}), a second set of simulation was performed. Although the efficiency map is based on PT efficiency characteristic -Type 2 where maximum efficiency is around 0.35, the optimal speed trajectories generated by the optimizer maintained
the same fundamental shape.  
The optimizer produced a speed profile which uses 272.38$kWs$ and 44.40\% less than the one obtained for the typical traffic baseline (Fig. \ref{fig:mainICE type}). 
Based on the third map, see Fig. \ref{fig:mainHybrid type}, the optimizer uses 518.28$kWs$ meaning a saving of 41.65\%. Again, the optimizer generates curves utilizing the highest efficiency
operating points in the $\eta(\omega,\text{T})$ diagram of the drivetrain.

The results obtained (Fig. \ref{fig:mainSpeed-dependent}) show that the optimized speed profile consumes 93.19kWs while the typical traffic baseline uses 124.68kWs, a 25.26\% reduction. The
optimal speed profiles generated by the optimizer maintained
the same fundamental shape, with a slight change. A difference is that the vehicle is driven at a constant speed especially in 
long segments instead of 
accelerating and coasting consecutively. This occurs because the efficiencies are the same for a specific speed and different torques. In this model, due to the presence of a constant speed section in the optimal trajectories, we obtained smaller savings. \\
Based on the generated curves shown in Fig. \ref{fig:mainTwo high ridges}, we obtained savings of 45.56\% when compared to the typical traffic baseline. Note that, the optimizer dropped a little in the region between two high efficiency areas to avoid consuming high power in that area. 

Numerous simulations were performed for the different efficiency maps utilizing the vehicle parameters
listed in Table \ref{table:I}. The results can be seen in Table 2 to Table \ref{table:VI}.
It is well known
that, most of the “typical trajectory” operates at
very low efficiencies. Also, the optimal trajectory results in energy savings,
which in the simulations shown in this paper range from 25\%
to 78\%, depending on the efficiency map. This
shows the robustness of the optimization with respect to the vehicle powertrain efficiency map.\\
In addition, based on the results, greater savings
are obtained for long segments. This is due to the optimizer
avoiding low torque and speeds, while the typical trajectory
operates mostly in less efficient regions of the efficiency map. Also, vehicle type 2 has higher maximum acceleration and deceleration which can produce greater savings in all the cases using different efficiency maps.\\ Vehicle type 2 uses more energy in both typical and optimal speed trajectories than vehicle type 1 because it is heavier and has larger cross-sectional area. Moreover, the optimizer achieved less energy savings using fourth efficiency map, see Table \ref{table:V}, in comparison to other efficiency maps due to the presence of the constant speed segments in optimal trajectories, especially during long segments. \\
\begin{table}[H]
\resizebox{15.8cm}{!}{\begin{minipage}{\textwidth}
\caption{ENERGY CONSUMPTION USING PT EFFICIENCY CHARACTERISTIC -TYPE 1}
\begin{tabular}{|l|c|c|c|c|c|}
\hline
      &             &             &             &               &                                       \\
Vehicle&Average speed&Segment length&Typical energy&Optimal energy& Energy saved\\
       &   $(m/s)$&   $(m)$      &   $(kWs)$        &    $(kWs)$          &                \\
\hline
       &                                     &    210       &273.40    &   91.62&  66.49\%    \\\cline{3-6}
       &                                       &    350       &362.82    &   102.60&  71.72\%      \\\cline{3-6}
Vehicle type 1                        &     7        &   700       &659.03    &   162.19&  75.38\%          \\\cline{3-6}
       &                                     &   2100       &1705.06    &   401.25&   76.47\%         \\\cline{2-6}
              &                              10       &   2100       &2010.41    &   577.81&   71.26\%         \\\cline{1-6}
\cline{1-6}
\hline
\hline
       &                                       &   210        &394.25    &   105.02&  73.37\%          \\\cline{3-6}
&                                     &   350        &612.20    &   137.34&  77.57\%          \\\cline{3-6}
Vehicle type 2&                  7     &   700       &1001.40    &   231.02&  76.90\%            \\\cline{3-6}
       &                                      &   2100       &2101.57    &   500.58&   76.24\%         \\\cline{2-6}
              &                              10       &   2100       &2585.37    &   568.15&   75.01\%         \\\cline{3-6}
\cline{1-3}
\hline
\end{tabular}
\end{minipage}}
\label{table:II}
\end{table}

\begin{table}[H]
\resizebox{15.7cm}{!}{\begin{minipage}{\textwidth}
\caption{ENERGY CONSUMPTION USING PT EFFICIENCY CHARACTERISTIC -TYPE 2}
\begin{tabular}{|l|c|c|c|c|c|}
\hline
      &                          &             &               &                        &                \\
Vehicle&Average speed&Segment length&Typical energy&Optimal energy& Energy saved\\
       &   $(m/s)$&   $(m)$      &   $(kWs)$        &    $(kWs)$          &                \\
\hline
       &                                      &    210       &375.55    &   230.87&  38.53\%    \\\cline{3-6}
       &                                       &    350       &489.96    &   272.38&  44.40\%      \\\cline{3-6}
Vehicle type 1                 &  7        &   700       &1304.00    & 692.90  &  46.86\%          \\\cline{3-6}
       &                                      &   2100       &2104.87    & 1079.60  &   48.70\%         \\\cline{2-6}
              &                         10       &   2100       &2256.70    &  1249.22&  44.64\%         \\\cline{1-6}
\cline{1-6}
\hline
\hline
       &                                     &   210        &623.80    &   310.95&  50.15\%          \\\cline{3-6}
&                                   &   350        &968.75    &   448.12&  53.74\%          \\\cline{3-6}
Vehicle type 2&                  7     &   700       &1679.57    &   775.70&  53.81\%            \\\cline{3-6}
       &                              &   2100       &2780.80    &   1192.35&   57.12\%         \\\cline{2-6}
              &                 10       &   2100       &3310.62   &   1406.60&   57.51\%         \\\cline{3-6}
\cline{1-3}
\hline
\end{tabular}
\end{minipage}}
\label{table:III}
\end{table}

\begin{table}[H]
\resizebox{15.7cm}{!}{\begin{minipage}{\textwidth}
\caption{ENERGY CONSUMPTION USING PT EFFICIENCY CHARACTERISTIC -TYPE 3}
\begin{tabular}{|l|c|c|c|c|c|}
\hline
      &             &             &             &               &                                      \\
Vehicle&Average speed&Segment length&Typical energy&Optimal energy& Energy saved\\
       &  $(m/s)$&   $(m)$      &   $(kWs)$        &    $(kWs)$          &                \\
\hline
       &                                   &    210       &663.51    &   401.79&  39.53\%    \\\cline{3-6}
       &                                     &    350       &888.30    &   518.28&  41.65\%      \\\cline{3-6}
Vehicle type 1              &     7        &   700       &1490.84    & 748.57  &  49.78\%          \\\cline{3-6}
       &                                &   2100       &3878.93   & 1962.04  &   49.42\%         \\\cline{2-6}
              &                           10       &   2100       &4213.02    &   2213.48&  47.46\%         \\\cline{1-6}
\cline{1-6}
\hline
\hline
       &                                  &   210        &854.22    &   466.29&  45.41\%          \\\cline{3-6}
&                                 &   350        &1201.15    &   568.34&  52.68\%          \\\cline{3-6}
Vehicle type 2&            7     &   700       &1714.88    &   817.35&  52.34\%            \\\cline{3-6}
       &                          &   2100       &4561.20    &   2018.65&   55.74\%         \\\cline{2-6}
              &                  10       &   2100       &4955.32    &   2215.01&   55.30\%         \\\cline{3-6}
\cline{1-3}
\hline
\end{tabular}
\label{table:IV}
\end{minipage}}
\end{table}

\begin{table}[H]
\resizebox{15.7cm}{!}{\begin{minipage}{\textwidth}
\caption{ENERGY CONSUMPTION USING PT EFFICIENCY CHARACTERISTIC -TYPE 4}
\begin{tabular}{|l|c|c|c|c|c|}
\hline
      &                   &             &               &                        &                \\
Vehicle&Average speed&Segment length&Typical energy&Optimal energy& Energy saved\\
       &  $(m/s)$&   $(m)$      &   $(kWs)$        &    $(kWs)$          &                \\
\hline
       &                                  &    210       &101.58    &   76.98&  24.21\%    \\\cline{3-6}
       &                                     &    350       &124.68    &   93.19&  25.26\%      \\\cline{3-6}
Vehicle type 1                 &     7        &   700       &182.14    & 29.59  &  49.78\%          \\\cline{3-6}
       &                             &   2100       &285.30   & 196.03  &   31.29\%         \\\cline{2-6}
              &                      10       &   2100       &365.21   &   264.14&  27.67\%         \\\cline{1-6}
\cline{1-6}
\hline
\hline
       &                                &   210        &174.35    &   122.25&  29.88\%          \\\cline{3-6}
&                        &              350        &201.10    &   132.14&  34.29\%          \\\cline{3-6}
Vehicle type 2&                 7     &   700       &256.34    &   162.26&  36.70\%            \\\cline{3-6}
       &                     &   2100       &341.02    &   214.29&   37.16\%         \\\cline{2-6}
              &                 10       &   2100       &414.80    &   285.68&   31.12\%         \\\cline{3-6}
\cline{1-3}

\hline
\end{tabular}
\label{table:V}
\end{minipage}}
\end{table}

\begin{table}[H]
\resizebox{15.8cm}{!}{\begin{minipage}{\textwidth}
\caption{ENERGY CONSUMPTION USING PT EFFICIENCY CHARACTERISTIC -TYPE 5}
\begin{tabular}{|l|c|c|c|c|c|}
\hline
      &             &               &               &                        &                \\
Vehicle&Average speed&Segment length&Typical energy&Optimal energy& Energy saved\\
       &  $(m/s)$&   $(m)$      &   $(kWs)$        &    $(kWs)$          &                \\
\hline
                         &             &    210       &173.90    &   99.26&  42.92\%    \\\cline{3-6}
       &                            &    350       &239.79    &   130.54&  45.56\%      \\\cline{3-6}
Vehicle type 1          &     7        &   700       &402.44    & 199.24  &  50.41\%          \\\cline{3-6}
       &                        &   2100       &1150.26   & 569.56  &   50.48\%         \\\cline{2-6}
                     &      10       &   2100       &1160.01   &   580.15&  49.98\%         \\\cline{1-6}
\cline{1-6}
\hline
\hline
       &                         &   210        &210.24    &   106.40&  49.39\%          \\\cline{3-6}
&                     &   350        &281.02    &  136.95&  51.27\%          \\\cline{3-6}
Vehicle type 2&        7     &   700       &492.30    &   237.88&  51.68\%            \\\cline{3-6}
       &                 &   2100       &1365.25    &   602.03&   55.90\%         \\\cline{2-6}
              &      10       &   2100       &1384.21    &   632.17&   54.33\%         \\\cline{3-6}
\cline{1-3}

\hline
\end{tabular}
\label{table:VI}
\end{minipage}}
\end{table}

\section{Transient Analysis}
\begin{figure}[H]
\begin{minipage}{.32\linewidth}
\centering
\subfloat[]{\label{withoutpenelty:a}\includegraphics[scale=.28]{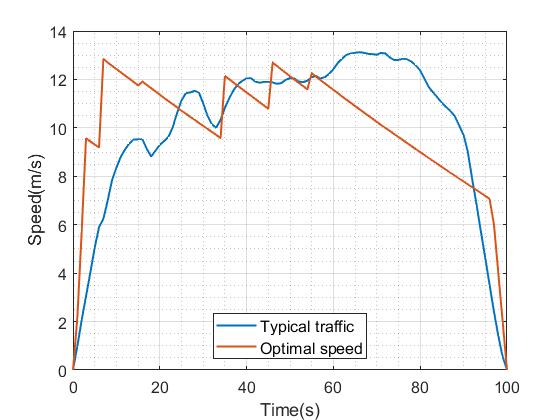}}
\end{minipage}
\begin{minipage}{.32\linewidth}
\centering
\subfloat[]{\label{withoutpenelty:b}\includegraphics[scale=.28]{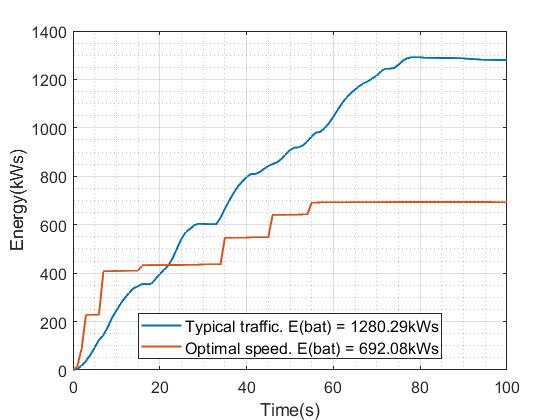}}
\end{minipage}
\begin{minipage}{.32\linewidth}
\centering
\subfloat[]{\label{withoutpenelty:c}\includegraphics[scale=.4]{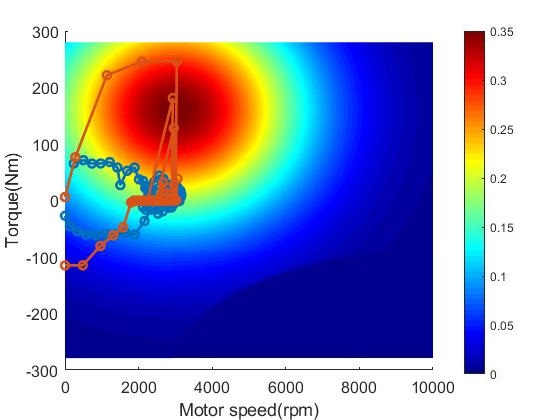}}
\end{minipage}
\caption[Two numerical solutions]{(a) Typical and optimal speed profiles, (b) cumulative energy consumption, (c) efficiency trajectories}
\label{fig:mainwithoutpenelty}
\end{figure}

\begin{figure}[H]
\begin{minipage}{.32\linewidth}
\centering
\subfloat[]{\label{withpenelty:a}\includegraphics[scale=.28]{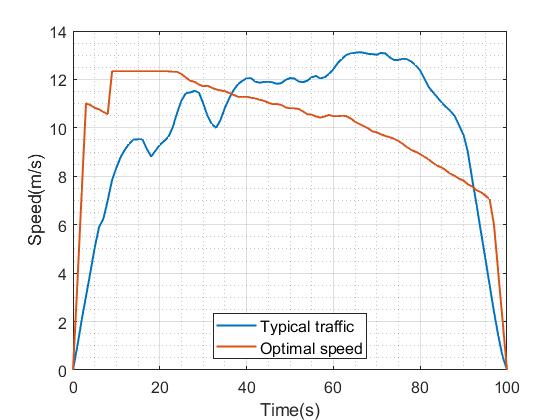}}
\end{minipage}
\begin{minipage}{.32\linewidth}
\centering
\subfloat[]{\label{withpenelty:b}\includegraphics[scale=.28]{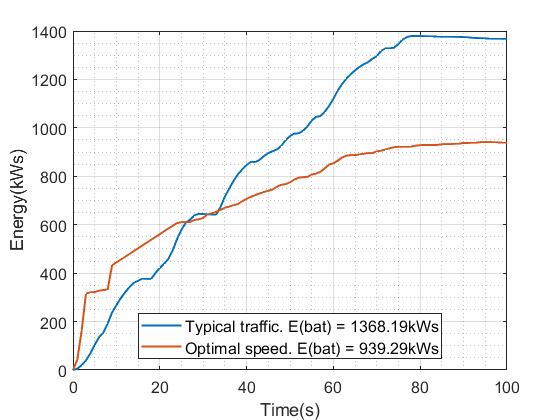}}
\end{minipage}
\begin{minipage}{.32\linewidth}
\centering
\subfloat[]{\label{withpenelty:c}\includegraphics[scale=.4]{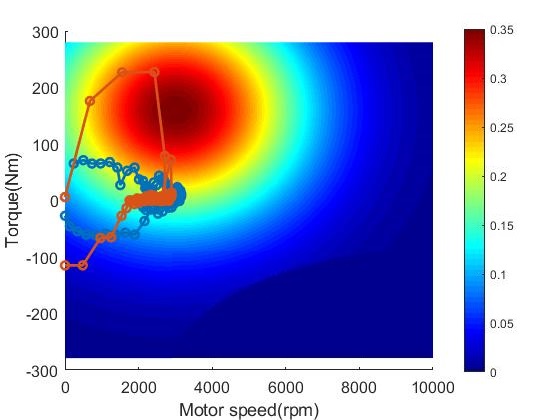}}
\end{minipage}
\caption[Two numerical solutions]{(a) Typical and optimal speed profiles, (b) cumulative energy consumption, (c) efficiency trajectories}
\label{fig:mainwithpenelty}
\end{figure}

In order to use a more accurate vehicle
model, we investigated an transient analysis where an increase in torque is penalized additionally which means that whenever the torque goes to wheels increases, an additional penalty (e.g. 10\%, 15\% and 30\%) for one second after that increasing will be added to the energy balance. The curves generated for a vehicle type 1 in a 1000m segment without and with adding 15\% additional penalty by using PT efficiency characteristic -Type 2 are shown in Fig. \ref{fig:mainwithoutpenelty} and Fig. \ref{fig:mainwithpenelty}, respectively.

It is shown that by adding a penalty to the energy balance, the optimizer tries to smoothly increase the torque instead of moving to higher torques suddenly. The results using different penalties can be seen in Table \ref{table:VII}. Noted that, as the penalty increases the energy saving decreases.   
\begin{table}[H]
\resizebox{13.1cm}{!}{\begin{minipage}{\textwidth}
\centering
\caption{ENERGY CONSUMPTION CONSIDERING ADDITIONAL PENALTY}
\begin{tabular}{|l|c|c|c|c|c|c|}
\hline
      &             &             &             &               &                        &                \\
Vehicle&Average speed&Segment length& Additional penalty &Typical energy&Optimal energy& Energy saved\\
       &  $(m/s)$&   $(m)$    &  &   $(kWs)$        &    $(kWs)$          &                \\
\hline
       &                         &             &    0\%       &1280.29   &   692.08&  45.94\%    \\\cline{4-7}
       &                          &             &    10\%       &1338.89    &   927.95&  30.69\%      \\\cline{4-7}
Vehicle type 1       &       10        &        1000    &  15\%       &1368.19    & 1002.52  &  26.72\%          \\\cline{4-7}
       &                         &             &   30\%       &1399.51  & 1291.35  &  7.72\%         \\\cline{1-7}

\hline
\end{tabular}
\label{table:VII}
\end{minipage}}
\end{table}

\section{Conclusion}
This paper presents the effect of the electric powertrain efficiency map on energy optimal speed trajectories. It is shown that the optimal speed trajectory is always approximately the same independent of the shape of the efficiency maps. However, the shape of the efficiency maps has a significant impact on the transportation energy savings (between 25\% and 78\%). \\
In addition, this paper examined the effects of different
vehicle parameters as well as segment lengths and average speeds. In all cases, the optimal
trajectory results in energy savings, depending on the constraints and parameters
chosen using different efficiency maps. This shows that the robustness of the optimizer with respect to efficiency maps and vehicle and infrastructure parameters. Vehicle type 2 can accelerate faster and has
greater savings in transportation energy, which is especially
pronounced in long segments. Also, energy savings
are particularly high during long distance segments.\\
Finally, the effect of adding a transient penalty to the energy balance on optimal profiles and savings using PT efficiency characteristic -Type 2 are presented. The shape of optimal trajectories remained the same as before for small penalties less than 10\%. However, by increasing the penalty from 10\% to 30\%, the optimizer tried to avoid going to high torque suddenly and it increased torque smoothly. It is also shown that, greater energy savings can obtained for smaller penalty percentages. The PT efficiency characteristic -Type 1 indicates that very large savings
are possible if there is a path with close to zero losses from zero rpm to the plateau region in the diagram.
This should inform the search for new PT technologies.

\bigskip
\section*{Authors}
\begin{minipage}[b]{21mm}
\includegraphics[width=20mm]{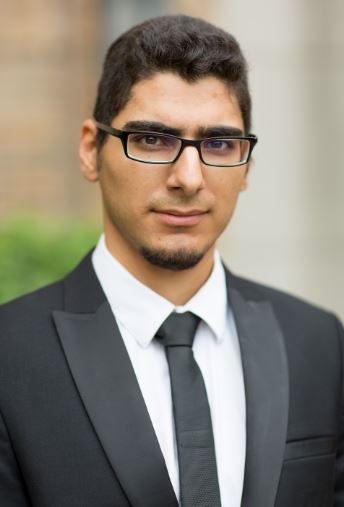}
\end{minipage}
\hfill
\begin{minipage}[b]{140mm}
\small
Arian Ahmadi received a B.Sc. in electrical engineering from IUST, Tehran, Iran, in 2017. He is currently a MS student in the Department of Electrical Engineering, University of Notre Dame, Notre Dame, IN, USA. His current researches focus on optimizing energy consumption for electric transportation, including estimation of vehicle parameters in networked vehicles.
\end{minipage}
\begin{minipage}[b]{21mm}
\includegraphics[width=20mm]{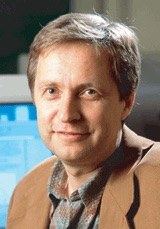}
\end{minipage}
\hfill
\begin{minipage}[b]{140mm}
\small
Peter H. Bauer received a Diploma from the
Technical University of Munich, Munich, Germany,
in 1984, and a Ph.D. degree from the University of
Miami, FL, USA, in 1987. He is a Professor in the
Department of Electrical Engineering, University of
Notre Dame, IN, USA. His research interests include
digital signal processing and control, sensor and actuator
networks, mobile wireless sensing, congestion
control, efficient and sustainable power generation
in transportation, and decentralized hybrid electric
power generation.
\end{minipage}

\begin{thebibliography}{99}
\small
\bibitem{Fnoor17}
Un-Noor, Fuad, et al. \emph{A comprehensive study of key electric vehicle
(EV) components, technologies, challenges, impacts, and future direction
of development.} Energies 10.8 (2017): 1217.

\bibitem{Zonggen14}
Z. Yi and P. H. Bauer. \emph{Energy Consumption Model and Charging Station Placement for Electric Vehicles.} Smartgreens. 2014

\bibitem{Zonggen2018}
Z. Yi and P. H. Bauer. \emph{Optimal stochastic eco-routing solutions for electric vehicles.} IEEE Transactions on Intelligent Transportation Systems 19.12 (2018): 3807-3817.

\bibitem{Hooker88}
J. N. Hooker.  \emph{Optimal driving for single-vehicle fuel economy.} Transportation Research Part A: General 22.3 (1988): 183-201.

\bibitem{Ahmadi}
A. Ahmadi, P. H. Bauer, Y. F. Huang, \emph{Estimating Environmental Parameters in Connected Electric Powertrains using Set-Membership Filtering.} Accepted in 2020 IEEE 91th Vehicular Technology Conference (VTC2020-Spring). IEEE, 2020.

\bibitem{Zonggen15}
Z. Yi and P. H. Bauer. \emph{Spatiotemporal energy demand models for electric vehicles.} IEEE Transactions on Vehicular Technology 65.3 (2015): 1030-1042.

\bibitem{Henriksson17}
M. Henriksson, O. Flärdh, and J. Mårtensson. \emph{Optimal speed trajectories under variations in the driving corridor.} IFAC-PapersOnLine 50.1 (2017): 12551-12556.

\bibitem{Eduardo19}
E. F. Mello and P. H. Bauer. \emph{Energy-Optimal Speed Trajectories Between Stops.} IEEE Transactions on Intelligent Transportation Systems (2019).

\bibitem{Eduardo20}
E. F. Mello and P. H. Bauer. \emph{Energy-optimal Speed Trajectories between Stops and Their Parameter Dependence.} VEHITS (2019).
 
\bibitem{Zonggen17}
Z. Yi and P. H. Bauer. \emph{Effects of environmental factors on electric vehicle energy consumption: a sensitivity analysis.} IET Electrical
Systems in Transportation, vol. 7, no. 1, pp. 3-13, 3 2017.


\bibitem{Zonggen18}
Z. Yi and P. H. Bauer. \emph{Energy Aware Driving: Optimal Electric
Vehicle Speed Profiles for Sustainability in Transportation.} IEEE
Transactions on Intelligent Transportation Systems, pp. 1-12, 2018.

\bibitem{Ehsani18}
M. Ehsani, Y. Gao, S. Longo, K. Ebrahimi. \emph{Modern Electric, Hybrid
Electric, and Fuel Cell Vehicles}. 3rd ed. Boca Raton, FL: CRC press,
2018.


\end{thebibliography}
\end{document}